# Sub-shot-noise-limit discrimination of on-off keyed coherent signals via a quantum receiver with a superconducting transition edge sensor

Kenji Tsujino,[1,2] Daiji Fukuda,[3] Go Fujii,[3,4] Shuichiro Inoue,[4] Mikio Fujiwara,[1] Masahiro Takeoka,[1] and Masahide Sasaki[1]

[1]*National Institute of Information and Communications Technology (NICT) 4-2-1 Nukui-kitamachi, Koganei, Tokyo, 184-8795, Japan*
[2]*Japan Science and Technology Agency, NEC Tsukuba Laboratories 34 Miyukigaoka, Tsukuba, Ibaraki305-8501, Japan*
[3]*National Institute of Advanced Industrial Science and Technology, 1-1-1 Umezono, Tsukuba, Ibaraki 305-8568, Japan*
[4]*Institute of Quantum Science, Nihon University, 1-8 Kanda-Surugadai, Chiyoda-ku, Tokyo 101-8308, Japan*

[*]*tsujino@qci.jst.go.jp*

**Abstract:** We demonstrate a sub-shot-noise-limit discrimination of on-off keyed coherent signals by an optimal displacement quantum receiver in which a superconducting transition edge sensor is installed. Use of a transition edge sensor and a fiber beam splitter realizes high total detection efficiency and high interference visibility of the receiver and the observed average error surpasses the shot-noise-limit in a wider range of the signal power. Our technique opens up a new technology for the sub-shot-noise-limit detection of coherent signals in optical communication channels.

## 1. Introduction

In optical communication systems, carriers are ideally prepared in coherent states. The conventional limit of discriminating coherent state signals is known as the shot-noise-limit which is achievable by measuring a physical quantity in which information is encoded, such as intensity or quadrature-amplitudes. This is, however, not the lowest error bound allowed by quantum mechanics.

The impossibility of a perfect discrimination of coherent states originates from the non-orthogonality of their wave functions. Quantum signal detection theory has been developed to investigate the best performance of discriminating non-orthogonal states and the mathematical expression of its quantum measurement [1]. For example, the optimal measurement of discriminating two coherent states with the minimum error is described by a projection measurement onto the quantum superposition of these states [2]. The realization of physical receivers acting as an optimal quantum measurement is important not only for fundamental physics but also for a direct application into optical communications such as space communications where the signals are sent through long-haul amplification-free channels. In particular, optimal quantum receiver beyond the shot-noise-limit for coherent signals (phase-shift keying, quadrature-amplitude modulation, etc.) is a promising device for the post coherent communication technology.

So far, several practical schemes of optimal or near-optimal quantum receivers have been theoretically suggested [3,4,5,6]. In particular, Kennedy proposed a simple near-optimal receiver to discriminate binary phase-shift keyed (BPSK) signals by use of a photon counter and a linear optical process (displacement operation) [6]. This was extended by Dolinar to an exactly optimal receiver by introducing a closed-loop feedback system [3] and its proof-of-principle experiment was recently demonstrated with an on-off keyed (OOK) signals, i.e. a coherent state and a vacuum [7]. In [7], an avalanche photodiode (APD) was used as a photon detector which allowed the real time electrical feedback enough faster than the optical pulse width (20 μs) while its system detection efficiency was relatively low (~35%). More recently, it was shown that Kennedy's idea can be generalized without using feedback by simply optimizing the displacement operation [5,8]. The proof-of-principle of this optimal displacement receiver was demonstrated for the BPSK signals with an APD [8]. Surprisingly, despite it's simple feedback-free configuration, the discrimination error performance shown in [8] was almost the same as that in [7]. However, due to the low detection efficiency of the system (~55%), it was not possible to go beyond the shot-noise-limit for the BPSK signals without compensating imperfections. It is therefore a necessary step to install highly efficient photon detectors into quantum receivers while keeping low dark counts.

In this paper, we report the first experiment of a quantum receiver in which a super conducting transition edge sensor (TES) is installed. TES is a photon-number resolving detector with very high detection efficiency (e.g. ~95% in [9]) and low noise and thus an attractive tool in quantum information technology. In fact, TES has been recently applied in quantum information experiments such as quantum key distribution [10,11] or quantum optics [12]. Here we apply a TES into quantum state discrimination scenario [1]. We use a titanium based superconducting TES [13] in the quantum receiver proposed in [5,8], which we call the optimal displacement receiver, and demonstrate the near-optimal discrimination of OOK signals below the shot-noise-limit. We achieve the total detection efficiency of the receiver as $70\pm3\%$ with the TES of $73\pm3\%$ detection efficiency at 853nm, which results the best error performance of the quantum receivers for binary coherent states reported so far. Our result is an important step toward practical quantum receivers overcoming the shot-noise-limit of conventional coherent optical communication systems.

## 2. Optimal displacement receiver

In this section we summarize a quantum mechanical formalism of the OOK signal detection problem and give a theoretical description of the optimal displacement receiver. In the OOK modulation, the signal is prepared in a coherent state $|\alpha\rangle$ or a vacuum $|0\rangle$ with the prior probabilities $p_\alpha$ and $p_0$, respectively. For simplicity, we assume $p_\alpha = p_0 = 1/2$. The task of the receiver here is to discriminate these states with errors as small as possible. The shot-noise-limit of the discrimination error is achieved by discriminating zero or non-zero photons via a direct on-off photon detector. An on-off detector is described by a set of operators $\{\hat{\Pi}_{ON} = \sum_{n=1}^{\infty} |n\rangle\langle n|, \hat{\Pi}_{OFF} = |0\rangle\langle 0|\}$, where $|n\rangle$ is an $n$-photon number state. The average error probability by the on-off detection, i.e. the shot-noise-limit, is calculated to be

$$p_e^{SNL} = \frac{1}{2}\left(\langle\alpha|\hat{\Pi}_{OFF}|\alpha\rangle + \langle 0|\hat{\Pi}_{ON}|0\rangle\right) = \frac{1}{2}e^{-|\alpha|^2} \qquad (1)$$

while the minimum error predicted by the quantum detection theory (quantum-limit) [1] is always lower than $p_e^{SNL}$ and given by

$$p_e^{QL} = \frac{1}{2}\left(1 - \sqrt{1 - e^{-|\alpha|^2}}\right). \qquad (2)$$

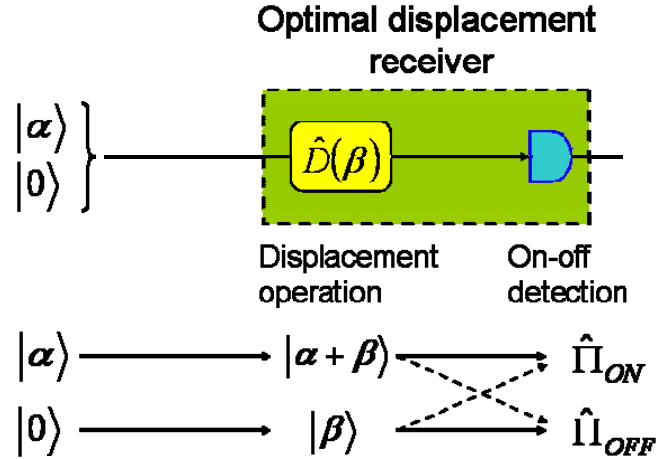

Figure 1: Schematic diagram of the optimal displacement receiver.

The schematic of the optimal displacement receiver is shown in Fig. 1. The signal is first shifted in phase space via a displacement operation as $\{|\alpha\rangle, |0\rangle\} \to \{|\alpha+\beta\rangle, |\beta\rangle\}$ which is realized by a transmissive beam splitter (with the transmittance $\tau \approx 1$) and a coherent local oscillator $|\beta/\sqrt{1-\tau}\rangle$. The average error probability of the state discrimination is then given by

$$p_e^{ODR} = \frac{1}{2}\left[1 - e^{-\nu - \eta|\beta|^2}\left\{1 - \exp\left[-\eta\left(\tau|\alpha|^2 + 2\xi\sqrt{\tau}\alpha\beta\right)\right]\right\}\right] \qquad (3)$$

where we have included realistic imperfections: the receiver's detection efficiency $\eta$, the dark counts $\nu$, and the mode match factor between the signal and the local oscillator $\xi$ [5]. The displacement parameter $\beta$ is chosen such that $p_e^{ODR}$ is minimized. Figure 2(a) shows an optimization of $\beta$. To illustrate the trade-off relation of each error, we plot the discrimination errors for the coherent state and the vacuum separately, which clearly suggests that the average error is minimized at certain $\beta$. The average error with optimal $\beta$ for various $|\alpha|^2$ is shown in Fig. 2(b). In ideal, the optimal displacement receiver surpasses the shot-noise-limit for any $|\alpha|^2$. We also plot the average errors for imperfect $\eta$ showing the drastic increase of errors with the decrease of $\eta$.

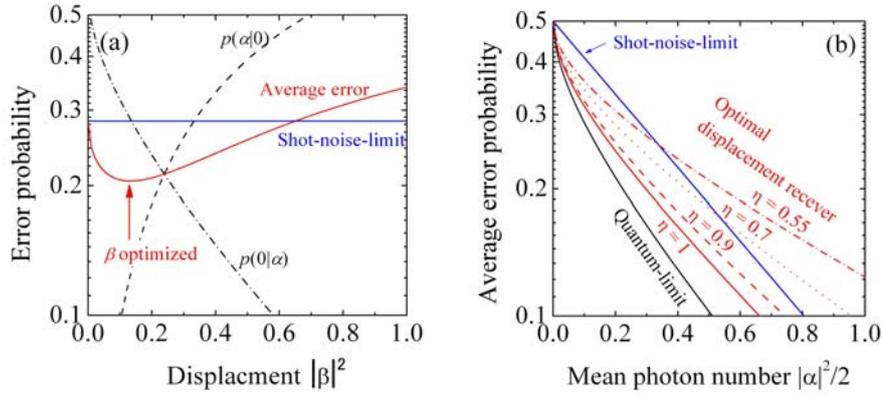

Figure 2: (a) Error probabilities as a function of displacement $\beta$. $p(\alpha|0)$ ($p(0|\alpha)$) is the probability of incorrectly guessing $|\alpha\rangle$ ($|0\rangle$). $|\alpha|^2$ =0.6, $\eta$ =1, $\nu$ =0, $\xi$ =1, and $\tau \to 1$. (b) Comparison of the average error probabilities of the shot-noise-limit (blue), the quantum-limit (black) and the optimal displacement receiver (red) with $\eta$ =0.55, 0.7, 0.9, and 1 (dotted-dashed, dotted, dashed, and solid lines). $\nu$ =0, $\xi$ =1, and $\tau \to 1$.

## 3. Transition edge sensor as an on-off detector

To make $\eta$ of the receiver as high as possible, we have used a TES as an on-off detector. Here we briefly describe our TES device and its operation as an on-off detector. TES is a calorimetric superconducting thermometer. The energy of absorbed photons at the TES film is sensed as the resistance increase and read out via a SQUID amplifier. Our TES consists of a $5 \times 5$ $\mu m^2$ titanium superconductor (30nm thickness) which is cooled down to 100mK whose quantum efficiency is $73 \pm 3\%$ at 853nm including the fiber-to-detector coupling loss. TES has the photon number resolving ability. Figure 3 shows an example of the pulse height distribution of the output voltage from our TES under 853nm laser pulse irradiation where the photon numbers are clearly distinguished. The energy resolution of the pulse height distribution is 0.4 eV. Details of the TES and its calibration method are found in [13]. An on-off detection with a TES is characterized by the detection efficiency and the dark counts. The on-off signal decision is done by setting a threshold in the pulse height distribution. The detection efficiency and the dark counts of the on-off detector are determined by the quantum efficiency of the device, the energy resolution, and the threshold. We choose the threshold as a half of the average pulse height of vacua and single-photons. Due to the high energy resolution, the detection efficiency of our detector is the same as the quantum efficiency of the device, i.e. 73±3%, while the dark counts of the TES itself are negligible.

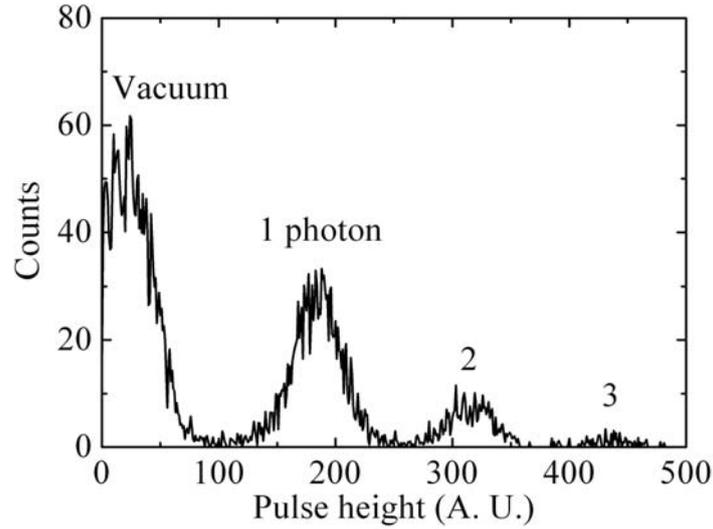

Figure 3: Histogram of the pulse height distribution of the TES. The TES is irradiated by 20ns laser pulses at 853nm. The peaks correspond to different numbers of the photons detected. The Gaussian-like noise at each peak is 0.28 times of the single-photon energy at 853nm.

## 4. Results

Figure 4 shows our experimental setup. Our source is a cw laser diode at 853 nm with a linewidth of less than 300 kHz. The laser light was pulsed by an EO amplitude modulator (AM0) with a pulse width of 20 ns and a repetition rate of 40 kHz. AM1 selects "on" (coherent state) and "off" (vacuum) signals. The signal pulses are transmitted through an optical fiber to the receiver. At the receiver, the signals are displaced by a controlled local oscillator via a fiber beam splitter (FBS) with a transmittance of 99% and directly guided into the TES to make the signal decision. The TES with a discriminator acts as a threshold detector and the error probabilities are calculated in a PC. The interferometer is actively locked before measurements. The interference visibility at the FBS is 98% corresponding to $\xi = 0.99$. The losses at the FBS and a splicing between the FBS and the fiber guided to the TES are estimated to be 3% in total, which results the total detection efficiency of the receiver as 70±3%. The total dark count of the receiver is 0.02 counts/pulse that are mainly caused by stray light.

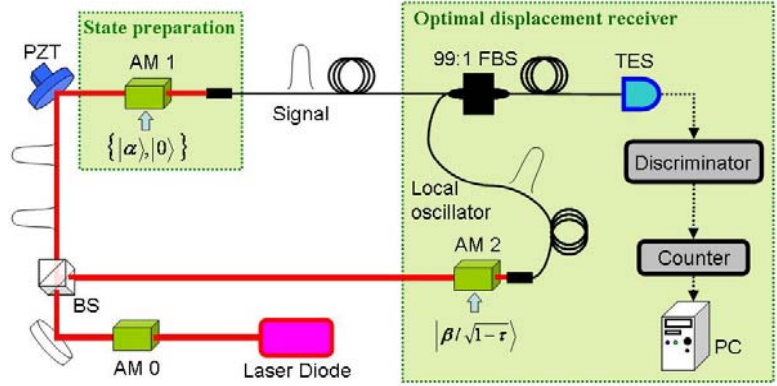

Figure 4: Experimental setup. AM: amplitude modulator, BS: beam splitter, FBS: fiber beam splitter, PZT: piezo-transducer, TES: titanium superconducting transition edge sensor.

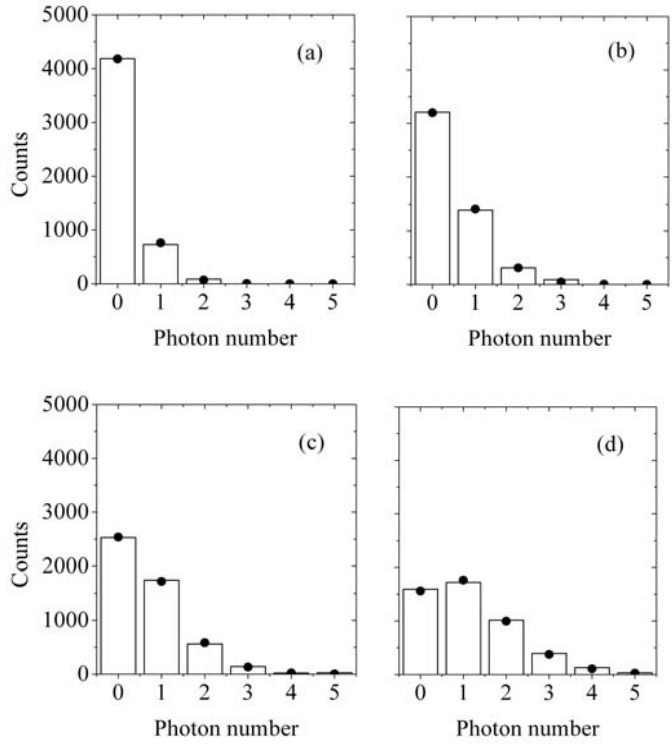

Figure 5: Photon number distribution of the signal states measured by the TES. The bars represent directly observed photon numbers and the circles are the Poissonian distribution with the average photon number of (a) $\bar{n} = 0.182$, (b) $\bar{n} = 0.441$, (c) $\bar{n} = 0.679$, and (d) $\bar{n} = 1.13$. Here $\bar{n}$ is experimentally estimated from $N_0/N_{\text{All}}$.

Photon number statistics of the signal pulses is characterized by the TES without a local oscillator. We compare directly observed photon number distributions from the TES and the Poissonian distribution

$$\text{Poisson}(n) = \frac{\bar{n}^n}{n!}\exp(-\bar{n}) \times N_{All} \qquad (4)$$

where $N_{All}$ is the total number of the events (=5000) and $\bar{n}$ is the average photon number derived as $\bar{n} = -\ln(N_0/N_{All})$ in which $N_0$ is the number of the zero-photon count events. In Fig. 5(a)-(d), the directly counted photon numbers (bars) and the Poissonian distributions (circles) are compared for several $|\alpha|^2$. We observe a nice coincidence and thus confirm that our signal source is classical noise-free. Calibration of the amplitude modulator for the signal state preparation (AM1) is also carried our by measuring the zero-photon count ratio.

For the OOK signal discrimination, we first measure the error probabilities for each signal by varying $\beta$ as shown in Fig. 6(a). Each plot with an error bar is obtained by repeating 8000 measurements and the minimization of the average error probability at certain $\beta$ is observed. It should be noted that the optimal $\beta$ experimentally observed does not always coincide with the theoretical prediction due to experimental imperfections. Figure 6(b) shows the experimental average error probabilities for various $|\alpha|^2$. The displacement $\beta$ is experimentally optimized at each $\alpha$. The squares are the experimental results and clearly surpass the (theoretical) shot-noise-limit (blue line) up to $|\alpha|^2/2 \approx 0.4$. Note that we do not make any compensation to the experimental data. The results almost agree with the theoretical line (red dotted line) in which the estimated imperfections are included. Some discrepancies between the theoretical curve and the data plots are due to the phase instability of the interference at the displacement FBS. The performance in our setup is also strictly limited by the non-unity detection efficiency. Loss of the efficiency in our TES is mainly due to the coupling loss between the fiber and the superconductor, which is an avoidable technical issue in future [9].

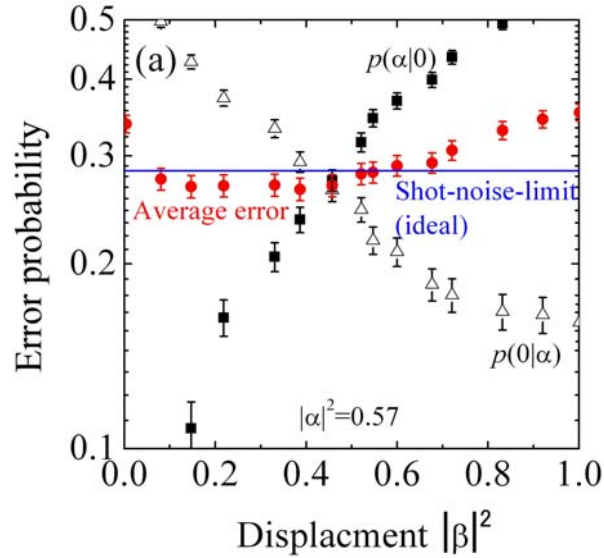

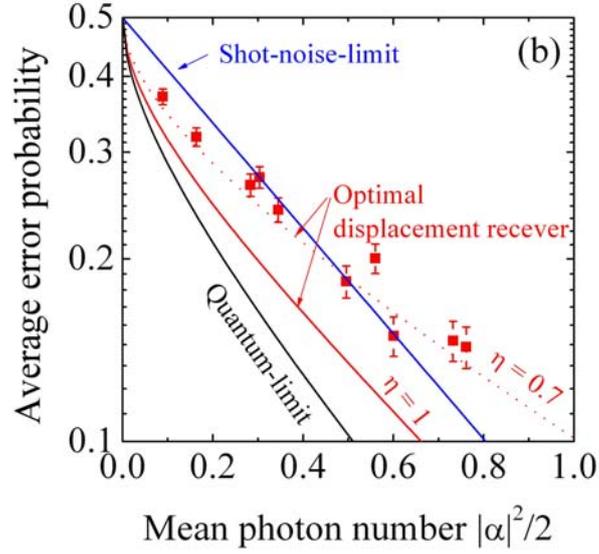

Figure 6: Experimental results. (a) Error probabilities p(α|0) (squares), p(0|α) (triangles), and their averages (circles) for various β. |α|²=0.57. (b) The average error probabilities with our optimal displacement receiver. The red dotted line indicates a theoretical curve of the receiver with η=0.7, ν=0.02, ξ=0.99, and τ=0.99. In both figures, the plots are obtained by 8000 measurements for each signal.

## 5. Summary

We have installed a superconducting transition edge sensor in the optimal displacement quantum receiver and demonstrated the sub-shot-noise discrimination of on-off keyed coherent states. Use of a TES and an FBS realized 70±3% of the total detection efficiency and 99% of the mode match factor, allowing us to surpass the shot-noise-limit in a wider range of the signal power. Our results showed a great potential of the TES to be used in quantum receivers for the discrimination of weak coherent signals. The non-unity detection efficiency of our TES is mainly due to the coupling loss between the fiber and the superconductor, which would be improved in future experiments.

One of the milestone in this direction is to go beyond the shot-noise-limit of discriminating phase-shift keyed or quadrature-amplitude modulated signals since these correspond to the theoretical limit of the detection sensitivity in the conventional optical communication technology (coherent communication). The biggest issue to be improved in future experiments is the total detection efficiency. For example, the numerical result [5] predicts that roughly speaking $\eta > 0.9$ and $\xi > 0.99$ are necessary to demonstrate the sub-shot-noise discrimination of BPSK signals with the optimal displacement receiver. However, our result and a rapid progress of the TES technologies strongly suggest its realization in near future.

### Acknowledgments

The authors would like to thank H. Takahashi and E. Sasaki for their technical supports.